# GRANULAR RHEOLOGY IN ZERO GRAVITY

G. Bossis[1], Y. Grasselli[2], O. Volkova[1]

[1] L.P.M.C. UMR 6622 – Université de Nice – Parc Valrose – 06108 Nice Cedex 2 (F)

[2] EAI Tech CERAM – Rue A. Einstein – BP 085 – 06902 Sophia Antipolis Cedex (F)

**Abstract**

We present an experimental investigation on the rheological behavior of model granular media made of nearly elastic spherical particles. The experiments are performed in a cylindrical Couette geometry and the experimental device is placed inside an airplane undergoing parabolic flights to cancel the effect of gravity. The corresponding curves, shear stress versus shear rate, are presented and a comparison with existing theories is proposed. The quadratic dependence on the shear rate is clearly shown and the behavior as a function of the solid volume fraction of particles exhibits a power law function. It is shown that theoretical predictions overestimate the experiments. We observe, at intermediate volume fractions, the formation of rings of particles regularly spaced along the height of the cell. The differences observed between experimental results and theoretical predictions are discussed and related to the structures formed in the granular medium submitted to the external shear.

## 1 - Introduction

The increase in interest in granular matter in recent years has been motivated by the existence of such materials in industrial and geological situations. Granular media exhibit amazing phenomena in both static and dynamic regimes. The shapes of heaps or the formation of arches in a static pile are still intriguing [1-2]. On the other hand, the rheological properties of a granular have shown several different behaviors depending on the flow range (quasistatic to rapid flows). A non exhaustive list includes non linear waves, inelastic collapse, segregation, convection rolls [3-5]… All these effects suggest that the local structure of the granular medium greatly influences its dynamical behavior. These aspects of granular flows have been investigated theoretically, by numerical simulations and experimentally [6-8].

Previous theories for rapid granular flows were derived from the classical solutions of the kinetic theory of gases. Nevertheless, important differences exist for a granular medium: the inelastic nature of collisions between particles and the fact that the velocity fluctuations can be of the same order of magnitude as the flow velocity. The determination of the granular temperature is obtained with the help of balance laws and taking into account the rate of dissipation during collisions [9]. The temperature is found to be proportional to the square of the shear rate, the square of a mean free path of the particles and inversely proportional to $1-e^2$, where e is the restitution coefficient of the grains of the granular medium which drives the energy dissipation phenomenon. Jenkins and Richman[10] have extended this theory by explicitly taking into account the anisotropy of the second moment of the velocity fluctuations. The density and velocity fields, as well as the velocity fluctuations, are then introduced in the balance laws of mass, momentum and energy to obtain the rheological description of a sheared granular medium. The boundaries, and specifically their roughness properties, also influence the rheology with the existence of a slip velocity [11] occurring at the moving walls of the cell.

In this paper we present, firstly, an experimental study of a sheared granular medium composed of iron spherical particles in a cylindrical Couette geometry. These experiments are performed in microgravity in order to be able to measure the viscosity of an homogeneous medium at high volume fraction without problems of convection and density fluctuations encountered in fluidized beds. Experiments focus on the dependence of the shear stress versus the shear rate and volume fraction (or density) of the medium. Finally, the experimental rheological curves are compared to existing theories and a discussion is proposed on the differences observed.

## 2 - Experiments

Experiments are performed in a cylindrical Couette geometry using two different types of model granular media composed of iron spherical particles with diameter $\sigma$ of 1mm and 2mm. The density of the particles $\rho_p$ is about 7500 kg/m$^3$ and the associated restitution coefficient e is close to 0.9. The cell has a height of 35.8mm, the inner cylinder a radius of 12.15mm and the outer one, a radius of 16.6mm. The number of layers of particles inside the shearing gap varies from 1 to 5 in our experimental situations. In this geometry, one can assume that the shear rate is almost constant. By changing the number of particles inside the cell, we are able to study the effect of the solid volume fraction $\nu$ (or density) on the rheology of the granular medium. A sketch of the experimental set-up is shown in figure 1. Iron particles of 1mm in diameter are glued to the inner cylinder to create a rough surface. The distance between these particles over the inner wall of the cell varies between 0.3 and 1 diameter. The inner cylinder is mounted on a rheometer which is computer controlled. A given angular velocity $\omega$ is applied to the inner cylinder and the rheometer measures the corresponding torque. This torque is then converted to a shear stress using the geometrical characteristics of the cell. In general, the relevant parameter of such studies is the ratio of the shear stress versus the normal stress but we only focus here on the shear stress behavior. The values of ω can be changed over a range varying from 0 up to 80rad/s, corresponding to a maximum shear rate of approximately 230 s$^{-1}$. The outer cylinder of the cell is made of glass to have a direct visualisation of the shearing process and to avoid electrostatic effects occurring with plastics.

To cancel the effect of gravity, experiments are performed inside an airplane undergoing parabolic flights. By parabola, we mean, in the following, the time interval during which the granular medium is no longer submitted to gravity. Each flight includes 30 successive parabolas, each lasting about 30s. The experimental determination of the rheological behavior of the granular medium is performed by measuring the torque exerted on the inner cylinder while changing the angular velocity over a given range. The torque recording sequence is started as soon as the apparent gravity drops to zero. However, the acceleration is monitored along the three directions and only the experimental data sets corresponding to nearly null fluctuations of the acceleration terms are kept.

To achieve experimental reproducibility, which is critical with granular media, several preliminary and test experiments have been realized. Setting the angular velocity to a constant value during a whole parabola does not exhibit significant fluctuations on the stress measurements. The standard deviation of the experimental values is less than 4%. Moreover, a recording sequence made by either increasing or decreasing the angular velocity over the same range of values gives practically the same behavior for the shear stress (fig. 2). Again, the average deviation on the recorded values is less than 5%. Finally, the rate of change of the angular velocity has no observable influence on the experimental results and is typically set between 0.5 and 2 rad/s². Nevertheless, to obtain sufficient experimental precision, the whole range of angular velocities cannot be investigated during a single parabola. That's why the experimental curves presented in this paper have been recorded over several successive parabolas. The continuity of the experimental data achieved in this way validates the experimental protocol.

Experimental curves of the shear stress τ versus the square of the shear rate $\dot{\gamma}^2$ for different solid volume fractions ν of particles are presented in figure 3 for the particles of 1mm and 2mm in diameter. As expected, the behavior is clearly quadratic in $\dot{\gamma}$ and one can also notice an increase with the volume fraction of particles.

## 3 - Analysis

Earlier work done by Bagnold [12] has shown that at rather high concentrations and shear rates, the corresponding stresses were function of the square of the applied shear rate. From a dimensional point of view, the shear stress is proportional to the density $\rho_p$ and square of the radius of the particles $\sigma^2/4$, the square of the shear rate $\dot{\gamma}^2$ and a tensor-valued function of the volume fraction, $\tau = \rho_p \sigma^2/4 \, f_{xy}(\nu)\dot{\gamma}^2$. Further experimental and numerical works [13-14] have shown that the function $f_{xy}(\nu)$ exhibited a "u-shaped" behavior with two vertical asymptotes, one for $\nu \rightarrow 0$ and the other for $\nu$ close to 0.6. In the range $0.1 < \nu < 0.5$, $f_{xy}(\nu)$ can be approximated by a power law, with an exponent depending on the theories [15]. A theoretical determination of the shear stress function $f_{xy}(\nu)$ has been derived in [10]. The dissipation of energy is determined by the frequency of collisions which is expressed in term of the pair distribution function between colliding particles. A local density is represented by the hard sphere pair distribution function at contact $g_0$. An expression has been proposed by Carnahan & Starling [16]:

$$g_0(\nu) = \frac{2-\nu}{2(1-\nu)^3} \qquad (1)$$

where $\nu$ is the solid volume fraction of the system. We note that the maximum compaction which can be achieved (RCP packing with a corresponding volume fraction of about 64%) should lead to an alternate expression of $g_0$ which diverges at this maximum volume fraction [7]. However, for values of $\nu$ up to 50%, the differences are small.

The velocity distribution function contains a perturbation to the Maxwellian distribution through the second and third moment of the fluctuating velocities. The associated coefficients are determined from the density, mean velocity, temperature and their spatial gradient.

In the continuation of the theoretical description, attention must be also taken on the boundary conditions. Rough parallel plates moving with a velocity U are considered with the granular medium sets in between. The geometry of the roughness surface of the boundaries play a non negligible effect on the granular flow. The main result of such consideration is the existence of a slip velocity V between the boundary and the closest layers of grains which tends to lower the stresses.

On this basis, Jenkins [17] and Richmann [18], have shown that for a homogeneous shear rate, the shear stress $\tau$ can be expressed in the form:

$$\tau = \rho_p f_{xy}(\nu)\sigma^2\dot{\gamma}^2\left(1-\frac{V}{U}\right)^2 \qquad (2)$$

where

$$f_{xy}(\nu) = \frac{1}{\sqrt{1-e}}\frac{\pi}{72}\nu\frac{1}{\sqrt{G}}\left[\frac{5}{16}\frac{1}{G}+1+\frac{4}{5}\left(1+\frac{12}{\pi}\right)G\right]^{3/2} \qquad (3)$$

and $G = \nu g_0(\nu)$ are function of the solid volume fraction $\nu$.

The expression of the slip velocity V is a complex function of $\nu$, $\sigma$, e, the number of layers of granular particles present in the sheared gap and the geometrical characteristics of the boundary roughness (a complete formulation of the slip velocity can be found in ref. [18], equations 31 & 32). Using equation (2), we can compare the theoretical shear stresses with the experimental ones. The roughness of the inner rotating wall is described by the ratio s/$\sigma$ where s is the average distance between glued particles. It is the only adjustable parameter between the theory and our experiments. Its value, sets to 0.4, is approximate because of the irregular distribution of glued particles but we have checked that in a reasonable range (0.3 to 0.8), it does not affect drastically the comparison with experimental results (the theory always overestimates the experiments).

Typical comparisons between experimental results and theory issued from equation (2) are presented in figure 4 for the particles of 1mm and the volume fractions of 14% and 42%. The discrepancy increases with increasing the volume fraction. On the other hand, it is noticeable that the theory predicts the right order of magnitude for the shear stress. The deviations could come from the type of structures observed in the cell during shearing and also from the fact that we have a small number of particle layers inside the gap of the cell. The slip velocity greatly increases with decreasing the number of layers. This is taken into account in equation 2 where the ratio of the thickness of the cell to the radius of the particles appears explicitly. But correlation lengths associated to clusters of particles should remain small compared to the gap between walls, which is not the case in our experiments, especially along the direction of flow.

Clustering effect may occur during the flow due to the inelastic nature of collisions between particles [19-20]. It depends on a characteristic length $L_* = \ell/\sqrt{1-e^2}$, where $\ell$ is the mean free path of particles $\ell = 1/(\pi n\sigma^2)$ (with n the number of particles by unit volume), compared to the size L of the system. If $L_*$ is larger than L, no cluster can exist but inhomogeneities can still be present. In our experimental situations, we do not observe any proof of the existence of "real" clusters but the particles seem to arrange themselves in annular structures (like rings of particles) regularly spaced in the vertical direction of the cell (fig. 5) and one could assume that these alignments are also present in the gap between the two cylinders of the cell. This arrangement of particles under shear seems to be more pronounced when the volume fraction is increased : we can guess it for the volume fraction of 28% and it is well formed at 42%. In this configuration, the momentum and energy transfers are preferentially performed between particles along the same line instead of in between particles of two different but adjacent lines, especially in the direction between the two walls of the cell. This should reduce the momentum transfer between the two walls of the cell. In this case, the slip, characterised by a slip velocity, would not only be determined by the geometrical properties of the roughness (as supposed in the theory) but could also be dependent on the structures present in the granular medium during shear. The fact that less momentum and energy could be transferred through particles in the direction perpendicular to the walls of the cell plays the same effect as a larger slip velocity but that would depend on the volume fraction. Actually we observe that the agreement between theory and experiment becomes worse as the volume fraction increases. Note that if one cancels the effect of slip velocity (setting V=0 in equation 2), the differences are even more important. This type of behavior and comparison between experiments and theory also act for the particles of 2mm in diameter. For this system, an additional effect may also be considered : the size difference between the particles of the medium and the particles glued on the inner cylinder to create a rough surface (two times smaller). This gives a lower effective roughness which may also increase the slip effect.

Experimentally, the dependence of the shear stress as a function of the volume fraction can be obtained from the slopes calculated on the curves of figure 3. Assuming all other parameters constant, the ratio $\tau/\dot{\gamma}^2$ is only a function of the volume fraction. The experimental values are presented in figure 6 and one can approximate the behavior obtained by a power law (dotted lines). The exponent calculated from this fit is, for the particles of 1mm and 2mm, respectively of 2.4 and 1.9, which also suggests a quadratic dependence on the solid volume fraction. In figure 6 are also reported the behavior of the ratio $\tau/\dot{\gamma}^2$ obtained using equation 2 (plain curves) for the two diameters of particles. Here too, the theory predicts a higher increase with the volume fraction. The values and the power law behaviors are quite different. Once again the existence of alignments at high volume fraction could be the explanation, but more systematic investigations have to be performed to conclude on this volume fraction dependence. Finally, we cannot, at this stage of the work, say more about the dependence on the size of the particles since we only used two different sizes.

4 - Conclusion

We have investigated the rheological behavior of model granular media made of nearly elastic spherical particles in Couette flow. The experiments have been performed in microgravity, in order to cancel the effects related to the weight of the particles and to be in an "ideal " situation to test the models of rapid shearing flows.

The shear stress exhibits a quadratic dependence versus the shear rate and a power law function versus the volume fraction as pointed out by previous experimental, numerical and theoretical works. Experimental results are compared to a theory issued from the kinetic theory of gases. The theoretical values of the shear stress are of the same order as, but still larger than the experimental values. The type of structures formed in the medium under shear, which tends to increase the effect of slip on the rotating cylinder, is the probable cause of these differences. When the volume fraction of the particles is increased, these structures tend to form concentric rings of particles in the direction of the flow which could introduce more slipping effects, thus reducing the shear stresses. More investigations have to be performed to clarify this point, in particular the use of a bidisperse medium which would prevent these structures to form allowing to better understand their effect on the rheology. New experiments focusing on the normal stress and on a systematic study on the influence of the size of the particles are underway.

**Acknowledgment**. This work has been supported by the National Center of Spatial Study (CNES) and we acknowledge for the possibility to use specially equipped aircraft for microgravity flights

**Captions of the figures**

Figure 1 : Sketch of the experimental setup used to record the shear stress of a model granular medium in a cylindrical Couette geometry. The CCD camera allows us to obtain a direct visualisation of the structures formed under shear. The experiments are performed in microgravity.

Figure 2 : Shear stress $\tau$ recorded over the same range of values of the shear rate $\dot\gamma$ on two different measurement sequences. The behavior obtained by either increasing (dotted curve) or decreasing (plain curve) the shear rate are similar.

Figure 3 : Shear stress $\tau$ as a function of the square of the shear rate $\dot\gamma^2$ for different solid volume fractions $\nu$. Iron spherical particles of 1mm (a) and 2mm (b) in diameter. The linear behaviors observed are a clear signature of the quadratic dependence on $\dot\gamma$.

Figure 4 : Comparison between experimental results (symbols) and theory retrieved from equation 2 (plain and dotted curves) for two volume fractions $\nu$. The overestimation of the theory in comparison to experiments increases with increasing $\nu$.

Figure 5 : Direct observations of the structures formed under shear for $\nu$=14%, $\nu$=28% and $\nu$=42% (particles of diameter 1mm). Left pictures : state of the granular without flow (in the presence of gravity). Right pictures : state of the granular with the presence of flow during a sequence of increase of the shear rate and with a zero gravity. The types of structures formed under shear are stable during a whole experimental recording sequence. An annular structure made of lines of particles in the direction of the flow appears at high volume fraction.

Figure 6 : Volume fraction dependence for the shear stress. The ratios $\tau/\dot\gamma^2$ retrieved from the experimental curves of figures 3 are plotted as a function of $\nu$ for the particles of diameter 1mm and 2mm. There is a power law dependence with exponents of 2.4 (1mm) and 1.9

(2mm). Experimental results : symbols and dotted lines. The plain curves represent the same behavior retrieved from equation 2 for the two diameters of particles.

# FIGURE 1

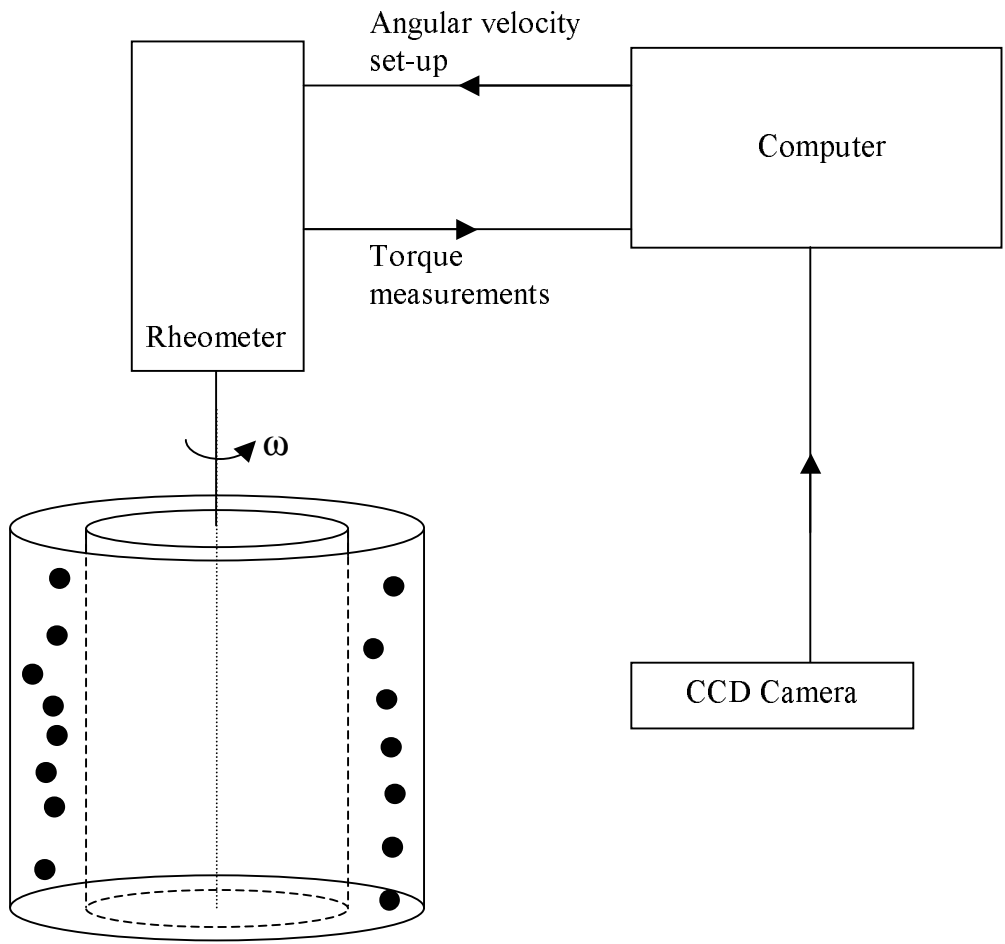

**FIGURE 2**

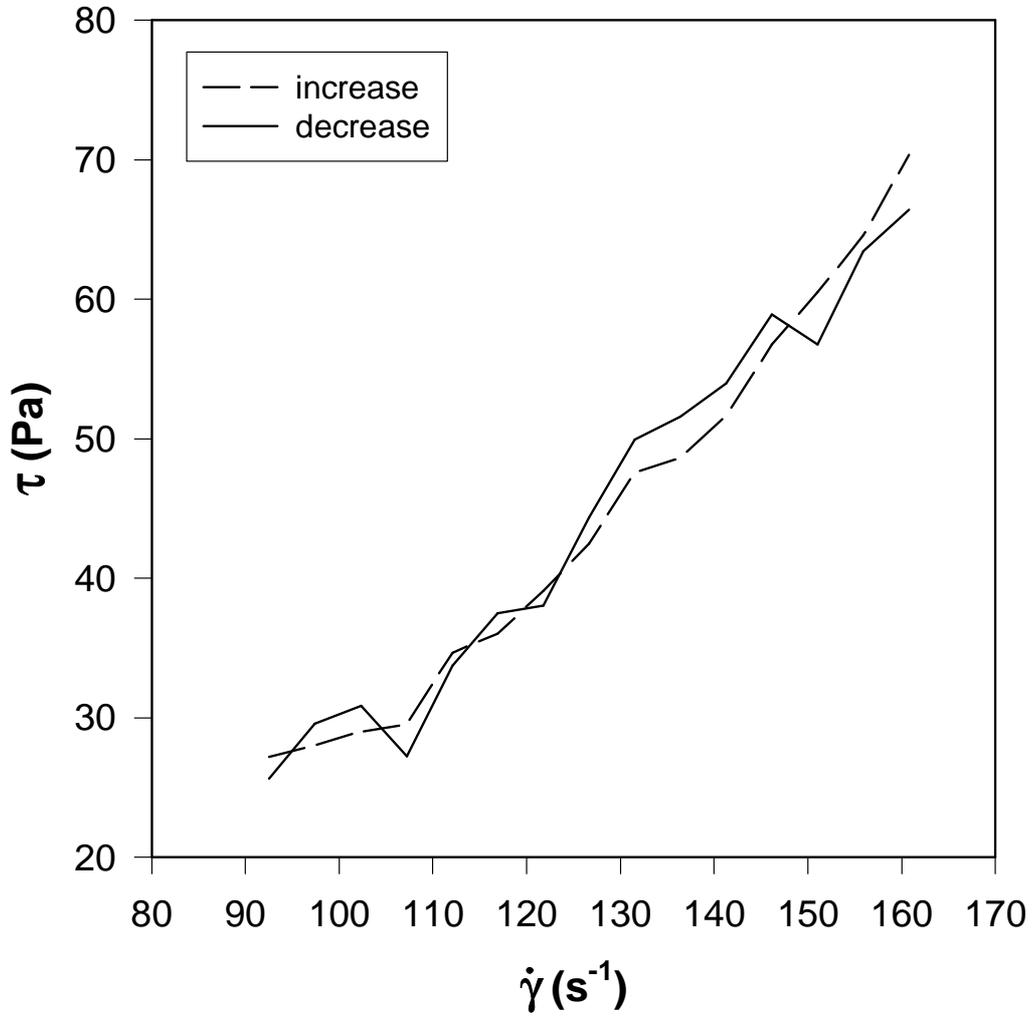

**FIGURE 3a**

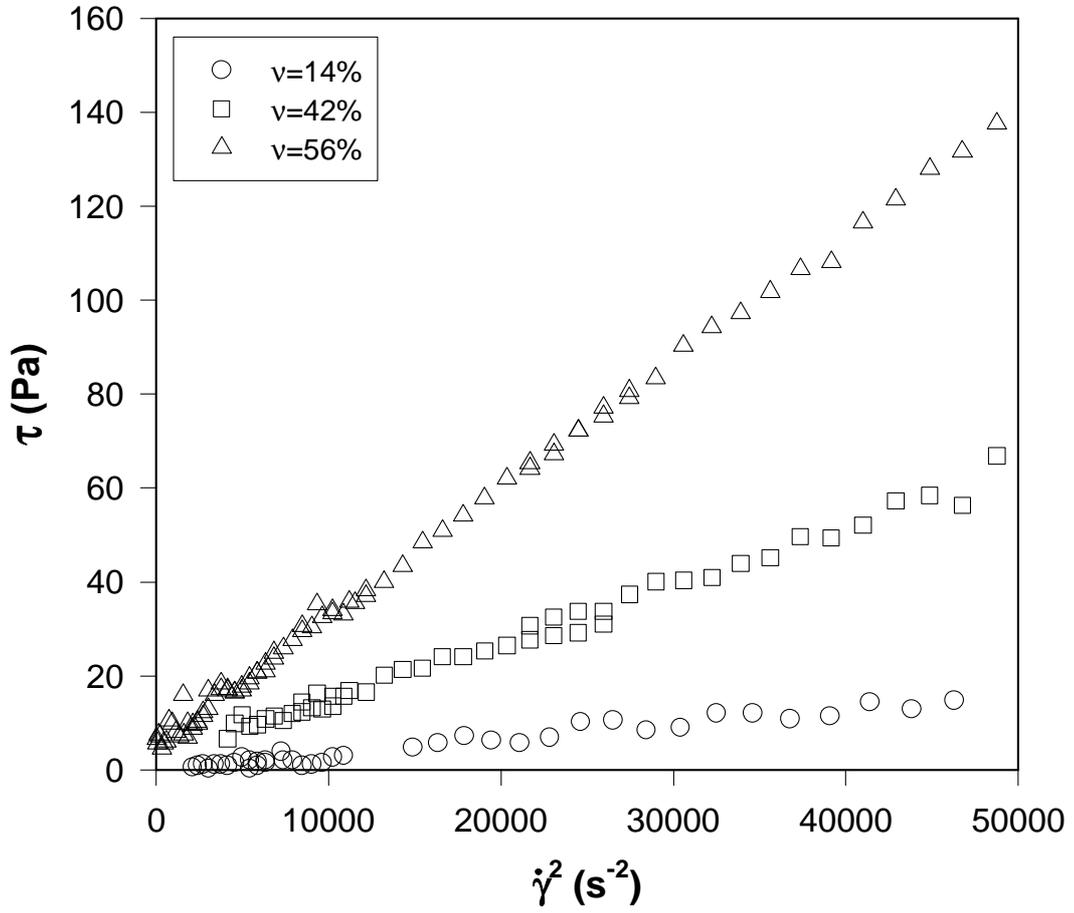

**FIGURE 3b**

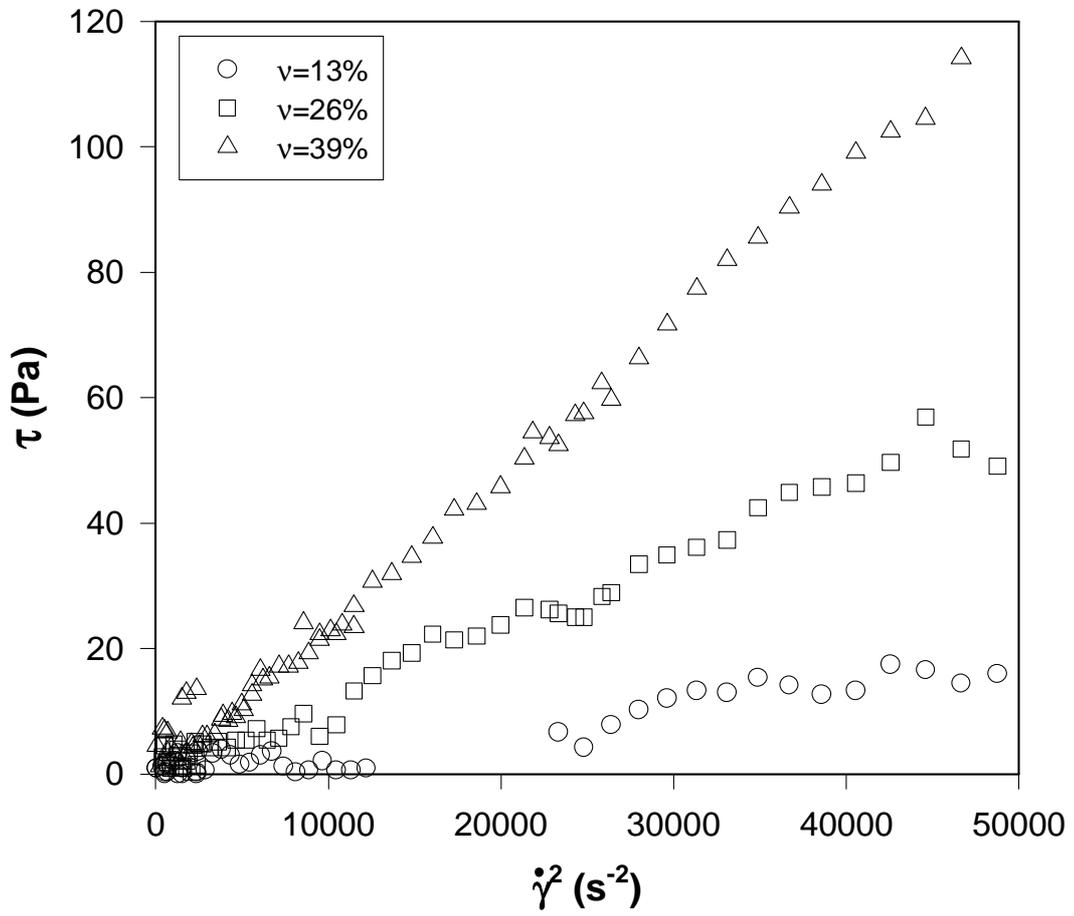

**FIGURE 4**

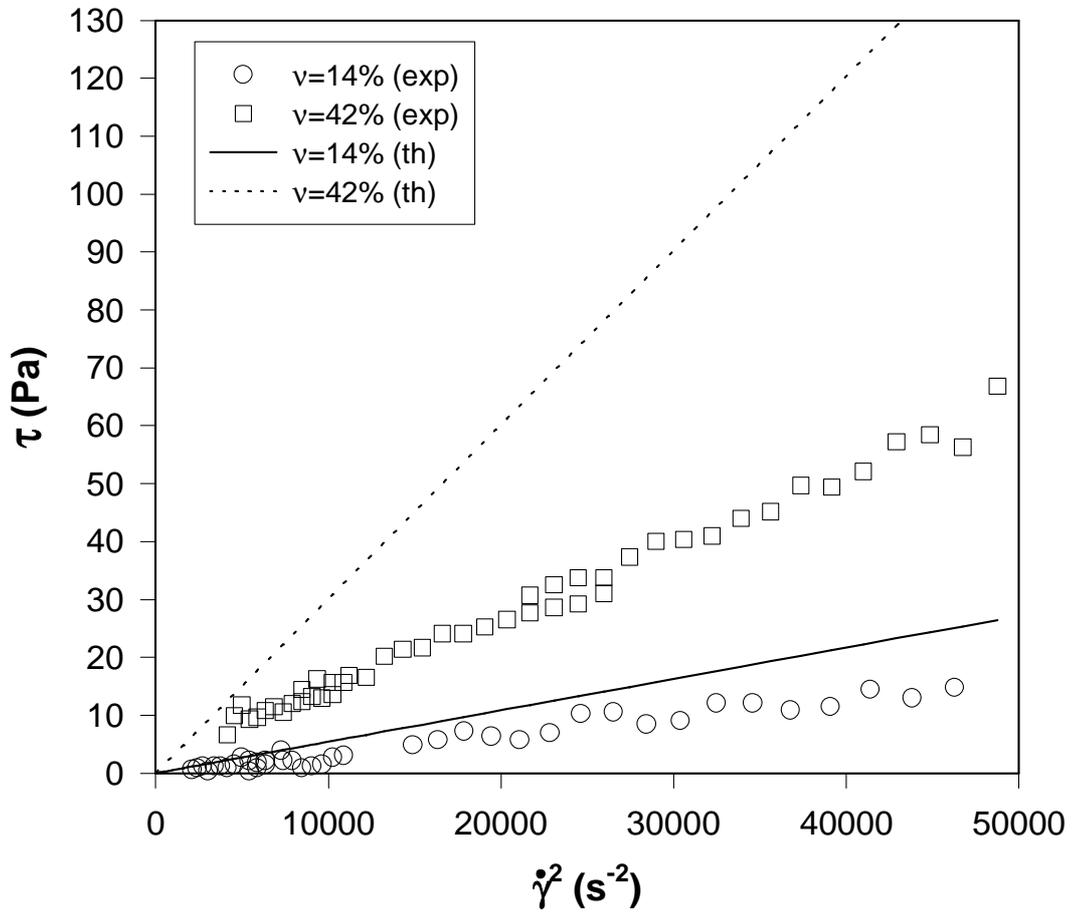

# FIGURE 5

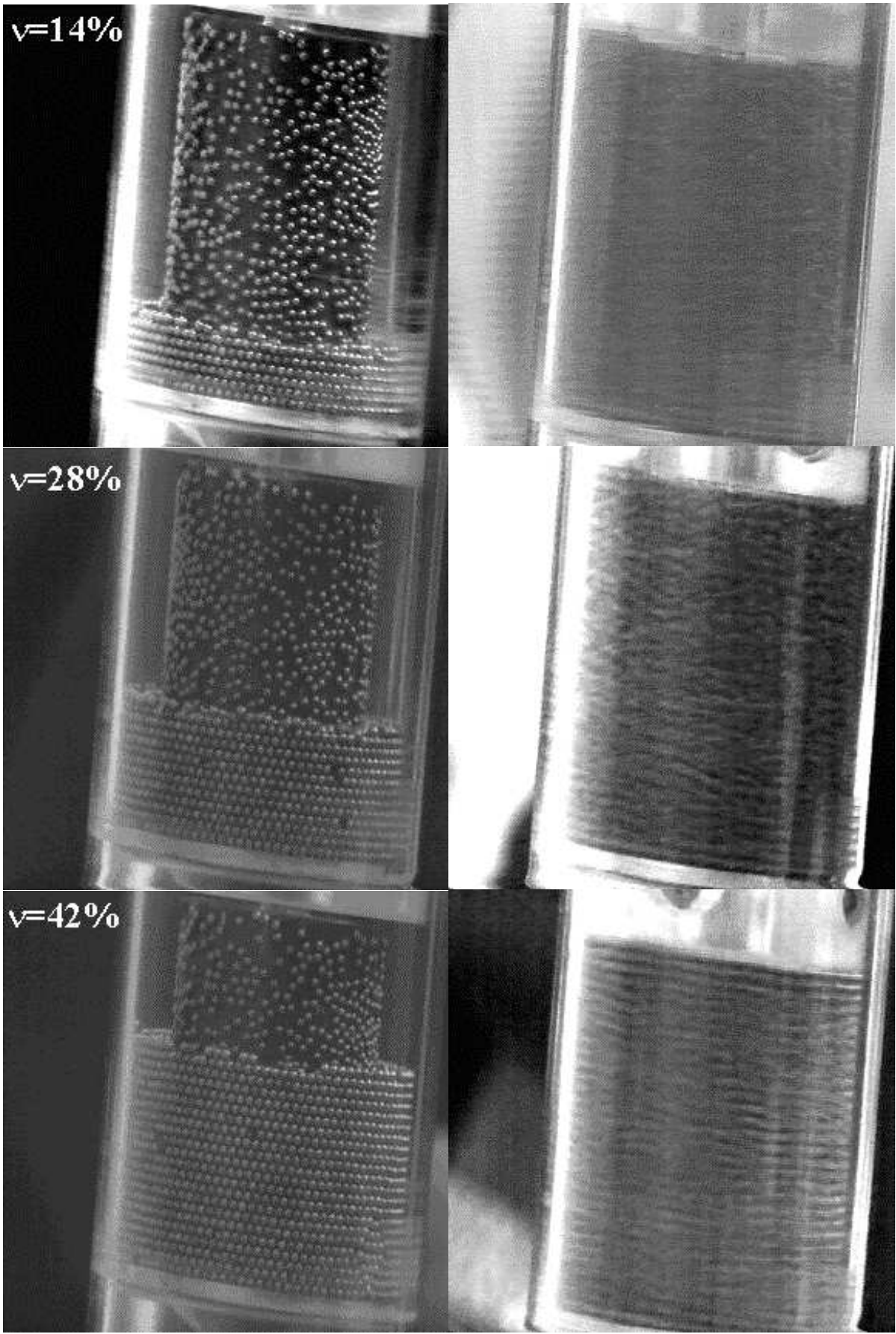

**FIGURE 6**

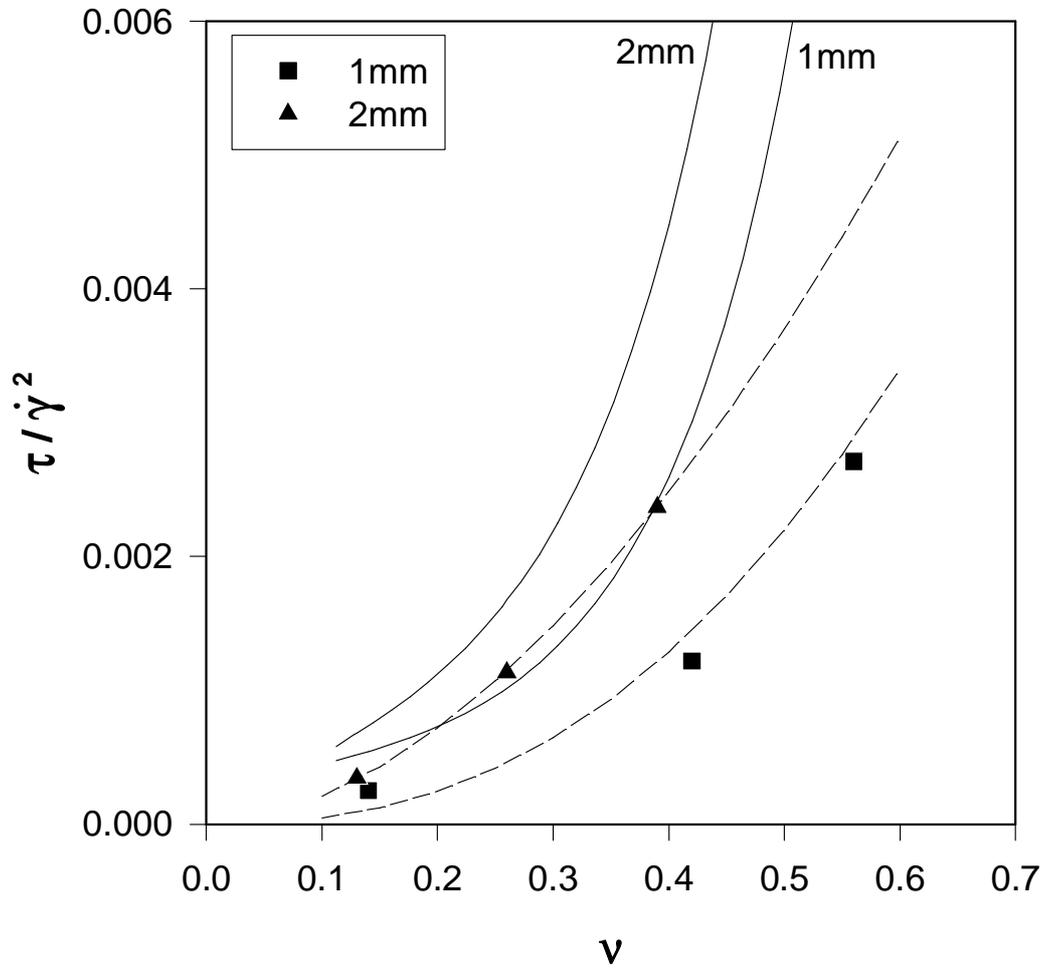